\documentclass[iop]{emulateapj}
\usepackage{apjfonts}
\usepackage{graphicx}
\usepackage{amsmath}

\slugcomment{Astrophysical Journal,accepted}

\begin{document}

%\submitted{The Astrophysical Journal Letter, submitted}   emul
%\vspace{1mm}


\title {CRAB NEBULA: FIVE-YEAR OBSERVATION WITH ARGO-YBJ}

\shorttitle{Crab Nebula by ARGO-YBJ} 


\shortauthors{Bartoli et al.} 
%\author{\centerline {{\it The ARGO-YBJ Collaboration: }}}
\author{B.~Bartoli$^{1,2}$,
 P.~Bernardini$^{3,4}$,
 X.J.~Bi$^{5}$,
 P.~Branchini$^{6}$,
 A.~Budano$^{6}$,
 P.~Camarri$^{7,8}$,
 Z.~Cao$^{5}$,
 R.~Cardarelli$^{8}$,
 S.~Catalanotti$^{1,2}$,
 S.Z.~Chen$^{5}$,
 T.L.~Chen$^{9}$,
 P.~Creti$^{4}$,
 S.W.~Cui$^{10}$,
 B.Z.~Dai$^{11}$,
 A.~D'Amone$^{3,4}$,
 Danzengluobu$^{9}$,
 I.~De~Mitri$^{3,4}$,
 B.~D'Ettorre Piazzoli$^{1,2}$,
 T.~Di~Girolamo$^{1,2}$,
 G.~Di~Sciascio$^{8}$,
 C.F.~Feng$^{12}$,
 Zhaoyang Feng$^{5}$,
 Zhenyong Feng$^{13}$,
 Q.B.~Gou$^{5}$,
 Y.Q.~Guo$^{5}$,
 H.H.~He$^{5}$,
 Haibing Hu$^{9}$,
 Hongbo Hu$^{5}$,
 M.~Iacovacci$^{1,2}$,
 R.~Iuppa$^{7,8}$,
 H.Y.~Jia$^{13}$,
 Labaciren$^{9}$,
 H.J.~Li$^{9}$,
 G.~Liguori$^{14,15}$,
 C.~Liu$^{5}$,
 J.~Liu$^{11}$,
 M.Y.~Liu$^{9}$,
 H.~Lu$^{5}$,
 L.L.~Ma$^{5}$,
 X.H.~Ma$^{5}$,
 G.~Mancarella$^{3,4}$,
 S.M.~Mari$^{6,16}$,
 G.~Marsella$^{3,4}$,
 D.~Martello$^{3,4}$,
 S.~Mastroianni$^{2}$,
 P.~Montini$^{6,16}$,
 C.C.~Ning$^{9}$,
 M.~Panareo$^{3,4}$,
 L.~Perrone$^{3,4}$,
 P.~Pistilli$^{6,16}$,
 F.~Ruggieri$^{6}$,
 P.~Salvini$^{15}$,
 R.~Santonico$^{7,8}$,
 P.R.~Shen$^{5}$,
 X.D.~Sheng$^{5}$,
 F.~Shi$^{5}$,
 A.~Surdo$^{4}$,
 Y.H.~Tan$^{5}$,
 P.~Vallania$^{17,18}$,
 S.~Vernetto$^{17,18,*}$,
 C.~Vigorito$^{18,19}$,
 H.~Wang$^{5}$,
 C.Y.~Wu$^{5}$,
 H.R.~Wu$^{5}$,
 L.~Xue$^{12}$,
 Q.Y.~Yang$^{11}$,
 X.C.~Yang$^{11}$,
 Z.G.~Yao$^{5}$,
 A.F.~Yuan$^{9}$,
 M.~Zha$^{5}$,
 H.M.~Zhang$^{5}$,
 L.~Zhang$^{11}$,
 X.Y.~Zhang$^{12}$,
 Y.~Zhang$^{5}$,
 J.~Zhao$^{5}$,
 Zhaxiciren$^{9}$,
 Zhaxisangzhu$^{9}$,
 X.X.~Zhou$^{13}$,
 F.R.~Zhu$^{13}$,
 Q.Q.~Zhu$^{5}$,
 G.~Zizzi$^{20}$\\ (The ARGO-YBJ Collaboration) and E.~Striani$^{21}$
}


\affil{$^{*}$Corresponding author: vernetto@to.infn.it\\
 $^{1}$Dipartimento di Fisica dell'Universit\`a di Napoli
                  ``Federico II'', Complesso Universitario di Monte
                  Sant'Angelo, via Cinthia, 80126 Napoli, Italy\\
 $^{2}$Istituto Nazionale di Fisica Nucleare, Sezione di
                  Napoli, Complesso Universitario di Monte
                  Sant'Angelo, via Cinthia, 80126 Napoli, Italy\\
 $^{3}$Dipartimento Matematica e Fisica "Ennio De Giorgi", 
                  Universit\`a del Salento,
                  via per Arnesano, 73100 Lecce, Italy\\
 $^{4}$Istituto Nazionale di Fisica Nucleare, Sezione di
                  Lecce, via per Arnesano, 73100 Lecce, Italy\\
 $^{5}$Key Laboratory of Particle Astrophysics, Institute
                  of High Energy Physics, Chinese Academy of Sciences,
                  P.O. Box 918, 100049 Beijing, P.R. China\\
 $^{6}$Istituto Nazionale di Fisica Nucleare, Sezione di
                  Roma Tre, via della Vasca Navale 84, 00146 Roma, Italy\\
 $^{7}$Dipartimento di Fisica dell'Universit\`a di Roma 
                  ``Tor Vergata'', via della Ricerca Scientifica 1, 
                  00133 Roma, Italy\\
 $^{8}$Istituto Nazionale di Fisica Nucleare, Sezione di
                  Roma Tor Vergata, via della Ricerca Scientifica 1,
                  00133 Roma, Italy\\
 $^{9}$Tibet University, 850000 Lhasa, Xizang, P.R. China\\
 $^{10}$Hebei Normal University, Shijiazhuang 050016,
                   Hebei, P.R. China\\
 $^{11}$Yunnan University, 2 North Cuihu Rd., 650091 Kunming,
                   Yunnan, P.R. China\\
 $^{12}$Shandong University, 250100 Jinan, Shandong, P.R. China\\
 $^{13}$Southwest Jiaotong University, 610031 Chengdu,
                   Sichuan, P.R. China\\
 $^{14}$Dipartimento di Fisica dell'Universit\`a di 
                   Pavia, via Bassi 6, 27100 Pavia, Italy\\
 $^{15}$Istituto Nazionale di Fisica Nucleare, Sezione di Pavia,
                   via Bassi 6, 27100 Pavia, Italy\\
 $^{16}$Dipartimento di Fisica dell'Universit\`a ``Roma Tre'',
                   via della Vasca Navale 84, 00146 Roma, Italy\\ 
 $^{17}$Osservatorio Astrofisico di Torino dell'Istituto Nazionale
                   di Astrofisica, via P. Giuria 1, 10125 Torino, Italy\\
 $^{18}$Istituto Nazionale di Fisica Nucleare,
                   Sezione di Torino, via P. Giuria 1, 10125 Torino, Italy\\
 $^{19}$Dipartimento di Fisica dell'Universit\`a di 
                   Torino, via P. Giuria 1, 10125 Torino, Italy\\
 $^{20}$Istituto Nazionale di Fisica Nucleare - CNAF, Viale
                   Berti-Pichat 6/2, 40127 Bologna, Italy\\
 $^{21}$Consorzio Interuniversitario Fisica Spaziale, 10133 Torino, Italy}

%\email{vernetto@to.infn.it}

\begin{abstract}

The ARGO-YBJ air shower detector monitored the Crab Nebula 
gamma ray emission from 2007 November to 2013 February.
The integrated signal, consisting of $\sim$3.3 $\times$ 10$^5$ events,
reached the statistical significance of 21.1 standard deviations.
The obtained energy spectrum in the energy range 0.3-20 TeV
can be described by a power law function
dN/dE = I$_0$ (E / 2 TeV)$^{-\alpha}$, with a flux normalization 
I$_0$ = (5.2 $\pm$ 0.2) $\times$ 10$^{-12}$ photons cm$^{-2}$ s$^{-1}$ TeV$^{-1}$
and $\alpha$ = 2.63 $\pm$ 0.05, corresponding to an integrated flux 
above 1 TeV of 1.97 $\times$ 10$^{-11}$ photons cm$^{-2}$ s$^{-1}$.
The systematic error is estimated to be less
that 30$\%$ for the flux normalization and 0.06 for the spectral index.
Assuming a power law spectrum with an exponential cutoff
dN/dE = I$_0$ (E / 2 TeV)$^{-\alpha}$ $\exp$ (-E / E$_{cut}$),
the lower limit of the cutoff energy E$_{cut}$ is 12 TeV,
at 90$\%$ confidence level.
Our extended dataset allows the study of
the TeV emission over long timescales.
Over five years, the light curve of the Crab Nebula
in 200-day bins is compatible with a steady emission with a probability of 
7.3 $\times$ 10$^{-2}$.
A correlated analysis with Fermi-LAT data over $\sim$4.5 years 
using the light curves of the two experiments 
gives a Pearson correlation coefficient $r$ = 0.56 $\pm$ 0.22. 
Concerning flux variations on timescales of days, 
a ``blind'' search for flares with a duration of 1-15 days gives
no excess with a significance higher than four standard deviations.
The average rate measured by ARGO-YBJ during the three most powerful flares
detected by Fermi-LAT is 205 $\pm$ 91 photons day$^{-1}$, 
consistent with the average value of 137 $\pm$ 10 day$^{-1}$.

\end{abstract}



\section{Introduction}

The Crab Nebula is the remnant of a supernova exploded in 1054 A.D.
at a distance of $\sim$2 kpc. 
It contains a 33 ms pulsar that powers a wind of relativistic particles.
The interactions of these particles with the remnant gas, photons and magnetic field
produce a non-thermal radiation extending from radio waves to TeV gamma rays
\citep[and references therein]{buh14}. 
Most of the emission is generally
attributed to synchrotron radiation of relativistic
electrons and positrons. The spectral energy distribution (SED) peaks
between optical and X-ray frequencies. A second component arises above 
$\sim$400 MeV,
interpreted as Inverse Compton (IC) of the same electrons scattering off 
synchrotron photons and CMB photons.

The Crab Nebula is one of the most luminous
sources of very high energy (VHE) gamma rays in the sky, 
and the first source to be detected at TeV energies \citep{wee89}.
Thanks to its high flux and apparent stability it is  considered
a reference source in gamma ray astronomy. Detected
by many experiments, both Cherenkov telescopes \citep{hegra04,hess06,magic08}
and air shower arrays \citep{tibet09,mila12}, 
the Crab Nebula was often used to check 
the detector performance, including sensitivity, 
pointing accuracy and angular resolution.

In 2010 September the AGILE satellite unexpectedly detected a strong flare
from the direction of the Crab Nebula at energies above 100 MeV. 
It lasted two days, with a maximum flux three times higher
than the average value \citep{agile1}, later confirmed by Fermi \citep{fermi1}. 
From then on, Fermi and AGILE reported some more flares, characterized by
a rapid increase and decay of the flux, typically lasting a few days.
The most impressive occurred in 2011 April, when the observed 
flux was $\sim$10 times higher than usual \citep{fermi2}.
The measured SED shows a new spectral component emerging during flares, 
peaking at high energies (up to hundreds MeVs in the 2011 April flare), 
attributed to a synchrotron emission
of a population of electrons accelerated up to energies of 
10$^{15}$ eV.
The Fermi-LAT data also show that
these sharp emission peaks are superimposed to long-lasting 
smoother modulations with timescales of weeks or months \citep{striani}.
The observed flux variations are attributed to the nebula, 
since the pulsar emission was found to be stable within 20$\%$ \citep{fermi2}.
However the origin of this activity is still unclear. 
In this scenario, 
observations at higher energies could provide precious information to 
understand the mechanisms responsible for this behaviour.

A preliminary analysis of the data recorded by the air shower detector
ARGO-YBJ during the flares,
showed an increase of the Crab flux at TeV energies
with a moderate statistical significance, in 3 out of 4 flares \citep{argo10,
ver_scineghe}.
These results have not been confirmed 
by Imaging Atmospheric Cherenkov Telescopes
(IACTs) because the Moon light hampered the observations.
However, sporadic and short measurements carried out during the first part 
of the 2010 September flare by MAGIC and VERITAS
show no evidence for a flux variability \citep{crab_cheren1,crab_cheren2}. 
The observations by VERITAS and HESS during a flare in 2013 March
(when ARGO-YBJ was already switched off),
report a counting rate consistent with the steady flux
\citep{veri13,hess13}.

In this paper we present a detailed analysis of the ARGO-YBJ data, 
carried out with a better reconstruction of the shower arrival direction,
obtained applying quality cuts on the events.
The study concerns not only the flaring episodes, 
but the whole Crab Nebula data set, consisting of more than five years of
observation.
The ARGO-YBJ layout and operation mode are presented in Section 2, with
a particular attention to the performance in gamma ray astronomy.
In Section 3 the analysis technique to extract the gamma ray signal
is outlined, followed by the results obtained with Crab Nebula data.
Section 4 reports the energy spectrum evaluation and discusses
systematic errors.
In Section 5 the analysis of the time behavior of the Crab Nebula signal
is presented, with a search for possible flares and rate variations
on different timescales. A time correlation
analysis with the Fermi-LAT data at energy E $>$ 100 MeV 
during $\sim$4.5 years is also reported.
Finally, Section 6 contains a summary of the results 
and concluding remarks.

\section{The ARGO-YBJ Experiment}

The ARGO-YBJ is a ``full coverage''  air shower detector located at 
the Yangbajing Cosmic Ray Laboratory (Tibet, P.R. China, 
longitude 90.5$^{\circ}$ East, latitude 30.1$^{\circ}$ North) 
at an altitude of 4300 m above sea level,
devoted to gamma ray astronomy at energies above $\sim$300 GeV 
and cosmic ray studies at energies above $\sim$1 TeV.

During its lifetime, from 2007 November to 2013 February, 
ARGO-YBJ monitored the gamma ray sky with an
integrated sensitivity ranging from 0.24 to $\sim$1 Crab Units \citep{sky13} 
and studied in detail the emission of the most luminous gamma ray sources
at energies above 300 GeV, namely the Crab Nebula,
MGRO J1908+06 \citep{mgro12}, HESS J1841-055 \citep{hessj13},
the Cygnus Region \citep{cyg12,cyg14}, and the blazars Mrk401 \citep{m421b,m421a} 
and Mrk501 \citep{m501}.

The detector consists of a 
$\sim$74 $\times$ 78 m$^2$ carpet made of a single layer of Resistive
Plate Chambers (RPCs) with $\sim$92$\%$ of active area, surrounded
by a partially instrumented ($\sim$20$\%$) area up to
$\sim$100 $\times$ 110 m$^2$. 
The apparatus has a modular structure,
with the basic data acquisition element being a cluster (5.7 $\times$ 7.6
m$^2$), made of 12 RPCs (2.85 $\times$ 1.23 m$^2$). 
Each RPC is read by 80 strips of 6.75 $\times$ 61.8 cm$^2$ (the
spatial pixels), logically organized in 10 independent pads of
55.6 $\times$ 61.8 cm$^2$ which are individually acquired and
represent the time pixels of the detector \citep{Aie06}.  
To extend the dynamical range 
up to PeV energies, each RPC is equipped with two large pads 
(139 $\times$ 123 cm$^2$) to collect the total charge developed 
by the particles hitting the detector \citep{ana12}.
The full experiment is made of 153 clusters (18360 pads), 
for a total active surface of $\sim$6600 m$^2$.

ARGO-YBJ operated in two independent
acquisition modes: the {\em shower mode} and the {\em scaler mode}
\citep{Aie08}. In this analysis we refer to the data recorded
from the digital read-out in shower mode.
In this mode, an electronic logic was implemented
to build an inclusive trigger, based on a time correlation 
between the pad signals, depending on their relative distance. 
In this way, all showers with a number of fired pads 
N$_{pad}\ge$ N$_{trig}$ in the central carpet in a time window 
of 420 ns generated the trigger.
This trigger worked with high efficiency down to N$_{trig}$ = 20,
keeping the rate of random coincidences negligible \citep{Alo04}.

The time of each fired pad in a window of 2 $\mu$sec around 
the trigger time and its location were recorded. 
To calibrate in time the 18360 pads, a
software procedure has been developed,
based on the Characteristic Plane method  \citep{He07}
that using the secondary particles of large vertical 
showers as calibration beams,
iteratively reduces the differences between the time
measurements and the time fit of the shower front  \citep{Aie09}.

The full detector was in stable data taking
with the trigger condition N$_{trig}$ = 20 
and an average  duty cycle $\sim$86$\%$. 
The trigger rate was $\sim$3.5 kHz with a dead time of 4$\%$.

The detector performance and capabilities in gamma ray astronomy have been
studied and improved through Monte Carlo simulations
describing the shower development in the atmosphere by using the 
CORSIKA code \citep{Hec98} and the detector response with a code based
on the GEANT package \citep{Gea93}.

\subsection{Field of View}

 \begin{figure}[t]
  \centering
  \includegraphics[width=0.45\textwidth]{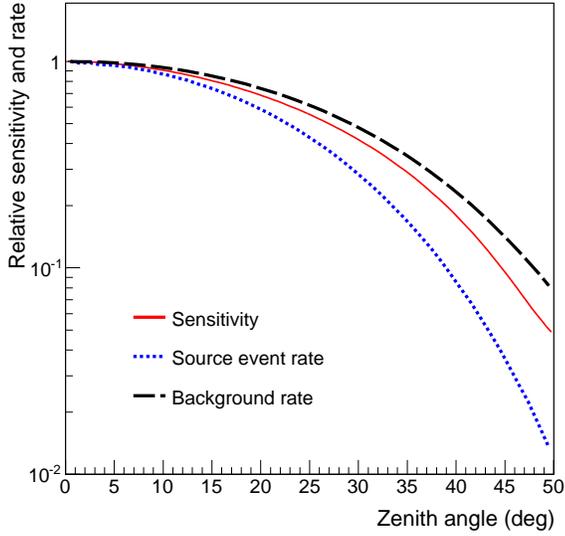}
  \caption{Sensitivity in the detector field of view.
Solid line: zenith angle dependence of the sensitivity to a Crab-like source. 
The sensitivity is normalized to a zenith angle $\theta$ = 0$^{\circ}$.
Dotted line: zenith angle dependence of the gamma ray event rate.
Dashed line: zenith angle dependence of the background rate; 
both rates are normalized to $\theta$ = 0$^{\circ}$.  }
  \label{sens}
  \vspace*{0.5cm}
 \end{figure}

One of the distinctive features of air shower arrays 
is the large field of view (FOV), in principle including 
the entire overhead sky.
Gamma ray sources cross the FOV with different paths according 
to their declinations.
The sensitivity is not uniform in the field of view.
Given a photon flux, the atmospheric absorption reduces the rate of showers
for increasing zenith angles. 
The cosmic ray background also decreases but more slowly, and the 
combination of the two rates determines the trend of the 
sensitivity as a function of the zenith angle.
Fig.\ref{sens} shows the event rate in ARGO-YBJ expected
from a Crab-like source as a function of the zenith angle $\theta$, 
normalized to the rate at $\theta$ = 0$^{\circ}$, compared to the background rate.
In the same figure the dependence of the detector sensitivity on 
$\theta$ is also reported. 
According to simulations, the sensitivity  at $\theta$ = 30$^{\circ}$ 
(45$^{\circ}$) is reduced by a factor $\sim$2 ($\sim$10)  
with respect to the sensitivity at $\theta$ = 0$^{\circ}$.

The capability to detect a given source depends on its path
in the field of view (determined by the source declination), 
and in particular on the amount of time that the
source lies at different zenith angles.
The maximum significance is for a declination $\delta_{max}$ = $\lambda$,
where $\lambda$ = 30.1$^{\circ}$ is the latitude of the detector.
Given a Crab-like source, the sensitivity decreases by less than 10$\%$ for 
declinations $\lvert \delta$-$\delta_{max}\rvert<$10$^{\circ}$, while it is reduced by a
factor $\sim$2 for declinations $\lvert \delta$-$\delta_{max}\rvert\sim$30$^{\circ}$.
The declination dependence is slightly stronger (weaker) for sources
with softer (harder) spectra with respect to the Crab Nebula
\citep {sky13}.  

%-------------------------------------------------------------------

 \begin{figure}[t]
  \centering
  \includegraphics[width=0.45\textwidth]{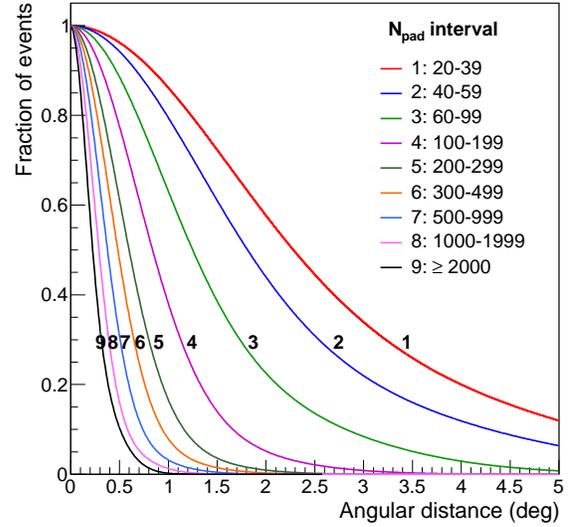}
  \caption{Angular resolution for different N$_{pad}$ intervals,
    according to simulations. The curves represent the fraction of events
    beyond the angular distance $d$ from the source, as a 
    function of $d$.}
  \label{psf}
  \vspace*{0.5cm}
 \end{figure}
%--------------------------------------------------------------------

At the ARGO-YBJ site, the Crab Nebula (declination $\delta$ = 22.01$^{\circ}$)
culminates at a zenith angle 
$\theta$ = 8.1$^{\circ}$ and lies at zenith angles $\theta$ $<$ 45$^{\circ}$
for 6.6 hours per sidereal day.
In general, following a source for a longer time per day increases the
signal significance, because of the increasing statistics, 
but since the signal to background ratio decreases at large zenith
angles, there is a maximum
zenith angle beyond which the significance begins to reduce. 
According to simulations, the maximum zenith angle for the Crab Nebula is 
$\sim$45$^{\circ}$.

\subsection{Angular Resolution}

%-------------------------------------------------------------------
  \begin{figure*}[t]
  \centering
  \includegraphics[width=0.80\textwidth]{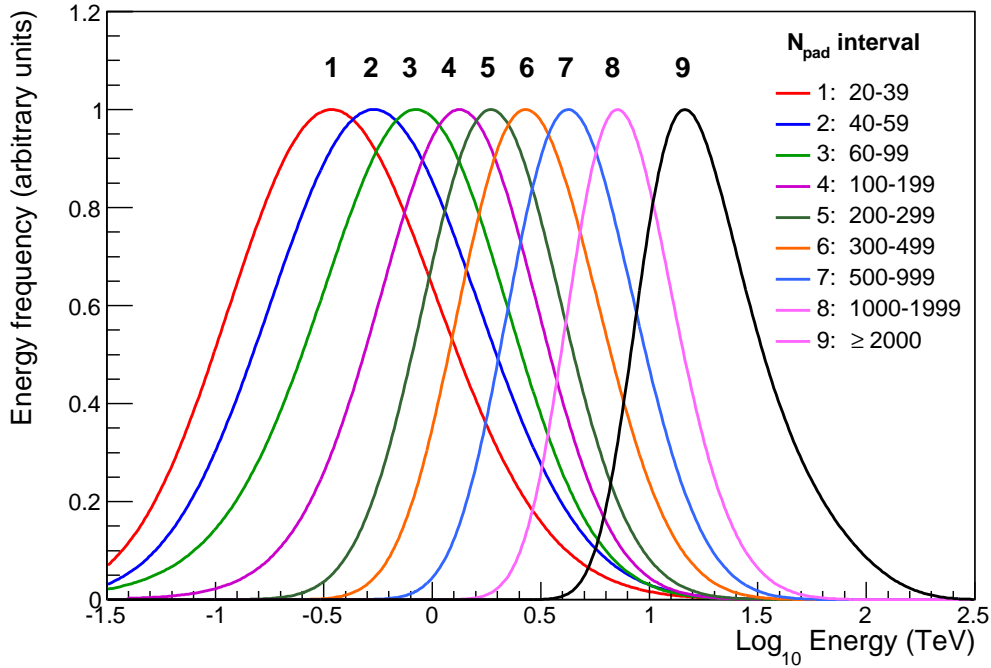}
  \caption{Normalized distribution of the primary gamma ray energy 
    for different  N$_{pad}$ intervals, for a Crab-like source.}
  \label{energy}
  \vspace*{0.5cm}
 \end{figure*}
%-------------------------------------------------------------------

The sensitivity needed to observe a gamma ray source is related to the angular 
resolution, which determines the amount of cosmic ray background.
We evaluate the shower arrival direction by fitting the shower front
with a conical shape centered on the shower core position,
to take into account the time delay
of secondary particles with respect to a flat front,
a delay that increases with the distance from the core.
We set this delay to 0.1 ns m$^{-1}$ \citep{Aie09}.

The high granularity of the detector allows the study of 
the shower profile  in great
detail and the accurate determination of the core position
by fitting the lateral density
distribution with a Nishimura-Kamata-Greisen-like function.
According to simulations,
the core position error depends on the number of hit pads
N$_{pad}$ and on the core distance from the detector center.
For gamma ray induced showers with a core distance less than 50 m,
the average core position error is
less than 8 (2) m for N$_{pad} \ge$ 100 (1000).

The point spread function (PSF) also depends on  N$_{pad}$, 
and for a given N$_{pad}$ value, it worsens as
the shower core distance from the detector center increases.
The angular resolution for showers induced by cosmic rays
has been checked by studying the Moon shadow, 
observed by ARGO-YBJ with a statistical significance 
of $\sim$9 standard deviations per month.
The shape of the shadow cast by the Moon on the cosmic ray flux
provides a measurement of the detector PSF.
This measurement has been found to be in excellent agreement with expectations, 
confirming the reliability of the simulation procedure \citep{DiS11}. 
 
The PSF for gamma ray showers is narrower than the cosmic ray one
by $\sim$30-40$\%$, due
to the better defined time profile of the showers.
To improve the angular resolution for gamma ray astronomy studies, 
quality cuts have been implemented,
by rejecting the events with a core distance  
larger than a given value $D_{cut}$ (depending on
N$_{pad}$) and with an average time spread of the particles
with respect to the fitted shower front exceeding 9 ns \citep{sky13}. 
The values of  $D_{cut}$ are given in Table \ref{table1}.
The fraction of gamma rays passing
the selection cuts depends on N$_{pad}$ averaging $\sim$80$\%$,
whereas the fraction of surviving background events is $\sim$76$\%$
for N$_{pad}$ $<$ 100 and $\sim$50$\%$ for  N$_{pad} \ge$ 100.
The selection also acts as a mild gamma/hadron discrimination 
for events with N$_{pad} \ge$ 100 (the sensitivity increases by a factor 
$\sim$1.1).

\begin{deluxetable}{ccccc}

\tabletypesize{\small}
\tablecaption{Characteristics of Crab Nebula simulated events.}
\tablewidth{0pt}
\tablehead{
\colhead {N$_{pad}$} & 
\colhead {$D_{cut}$\tablenotemark{a}} & 
\colhead {Core position} & 
\colhead {$R_{39}$\tablenotemark{c}} & 
\colhead {Median energy} \\ 
\colhead {} & 
\colhead {(m)} & 
\colhead {error\tablenotemark{b} (m)} & 
\colhead {(deg)} & 
\colhead {(TeV)} 
}
\startdata
20-39     &  no limits & 37  &  1.88 & 0.34  \\ 
40-59     &  no limits & 28  &  1.50 & 0.53  \\ 
60-99     &  90        & 12  &  1.04 & 0.79  \\ 
100-199   &  70        & 6.8 &  0.70 & 1.3   \\ 
200-299   &  60        & 4.2 &  0.50 & 2.1   \\ 
300-499   &  60        & 3.3 &  0.41 & 3.1   \\ 
500-999   &  40        & 2.3 &  0.32 & 4.8   \\ 
1000-1999 &  30        & 1.6 &  0.24 & 8.1   \\ 
$\ge$ 2000 &  30        & 1.0 &  0.19 & 17.7  
\enddata

\tablecomments{
\\$^a$ Maximum distance of the shower core from the detector center 
beyond which the events are rejected.
\\$^b$  Distance between the true and the reconstructed cores
containing 68$\%$ of the events.
\\$^c$ Angular resolution, defined as the 39$\%$ containment radius.
}

\label{table1}
\end{deluxetable}

%------------------------------------------------------------------

The arrival directions of the selected showers are also corrected
for the systematic error due to the partial sampling of the shower front
when the core is close to the edge of the detector \citep{eck}.
This systematic error is related to the angle between the vector 
``shower core-detector center'' and the shower arrival direction. 
For events with N$_{pad} \ge$ 100, for which the core position is
determined with more accuracy, the error can be considerably reduced. 

These selections and corrections shrink the PSF by a factor ranging from
$\sim$1.1 for events with N$_{pad}$ = 20-39, 
up to $\sim$2, for N$_{pad} \ge$ 1000. 
The PSFs obtained by simulating the Crab Nebula
along its daily path up to $\theta$ = 45$^{\circ}$
are shown in Fig.\ref{psf} for different intervals of N$_{pad}$.

To describe the PSFs analitically, that for small values of N$_{pad}$
cannot be simply fitted by a two-dimensional Gaussian function, 
the simulated distributions
have been fitted with a linear combination of two Gaussians.
In general, when the PSF is described by a single 
Gaussian (F($r$) = 1/(2$\pi$$\sigma^2$) 
 $\exp $ (-$r^2$/$\sigma^2$), where $r$ is the angular distance
from the source position), the value of the root mean square $\sigma$ 
is commonly defined as the ``angular resolution''. In this case the fraction
of events within 1 $\sigma$ is 39$\%$.
For our PSFs, the value of the  
39$\%$ containment radius $R_{39}$ ranges from
0.19$^{\circ}$ for  N$_{pad} \ge$ 2000 to 1.9$^{\circ}$ for  N$_{pad}$ = 20-39.
Table \ref{table1} reports the values of  $R_{39}$ for
different  N$_{pad}$ intervals, together with the
core position error, after quality cuts,
as obtained by simulating the source during the daily path in the
ARGO-YBJ field of view.

%-------------------------------------------------------------------
 \begin{figure}[t]
  \centering
  \includegraphics[width=0.45\textwidth]{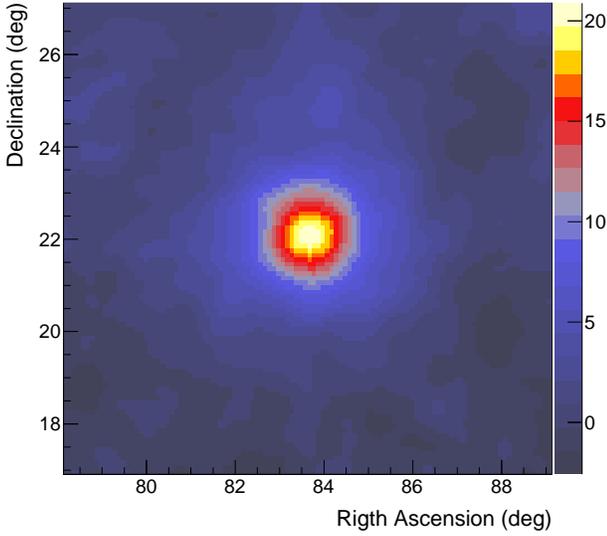}
  \caption{Significance map of the Crab Nebula region, after smoothing 
and background subtraction.}
  \label{map}
  \vspace*{0.5cm}
 \end{figure}
%-------------------------------------------------------------------

%------------------------------------------------------------------------
 \begin{figure}[t]
  \centering
  \includegraphics[width=0.45\textwidth]{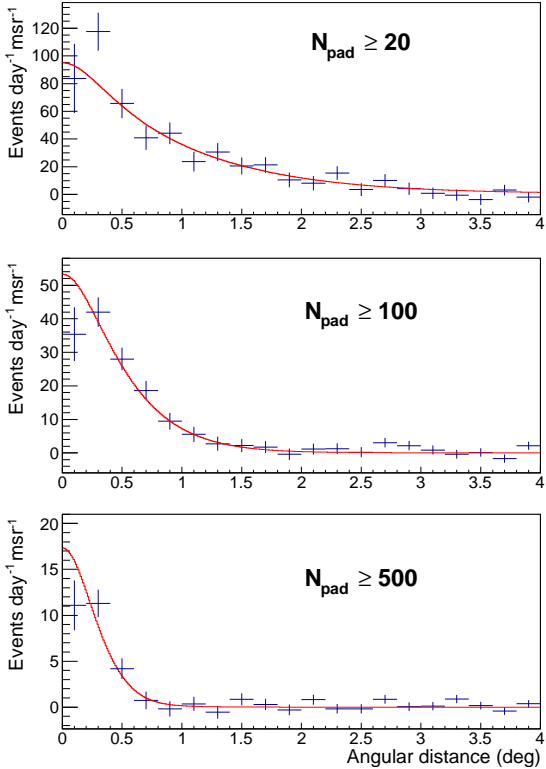}
  \caption{Comparison of experimental and simulated PSFs: 
event rate per solid angle as a function of the angular 
distance from the source position, 
for events with N$_{pad} \ge$ 20, 100 and 500.
The solid lines are the results of the simulation.}
  \label{ang_res}
  \vspace*{0.5cm}
 \end{figure}

%------------------------------------------------------------------------

\subsection{Energy Measurement}

The number of hit pads N$_{pad}$ is the observable
related to the primary energy that is used to infer the source spectrum.
In general, the  number of particles at ground level is not a very
accurate estimator of the primary energy of the single event, 
due to the large fluctuations in the 
shower development in the atmosphere.
Moreover, for a given shower, the number of particles detected
in a finite area detector like ARGO-YBJ 
depends on the position of the shower core with respect 
to the detector center; for
small showers this is especially poorly determined.

The relation between N$_{pad}$ and the 
primary gamma ray energy of showers surviving the selection cuts
is illustrated in Fig.\ref{energy}, where the corresponding 
primary energy distributions for different N$_{pad}$ intervals 
are reported, as obtained by simulating
a Crab-like source with a power law spectrum with index -2.63.
The distributions are broad, with extended overlapping regions,
spanning over more than one order of magnitude for small N$_{pad}$ values.
The median energies for different  N$_{pad}$ intervals are given
in Table \ref{table1}. They range from 340 GeV for events with 
N$_{pad}$ = 20-39, to $\sim$18 TeV for N$_{pad} \ge$ 2000.

Since the variable N$_{pad}$ does not allow the accurate measurement of the 
primary energy of the single event, the energy spectrum
is evaluated by studying the global distribution of N$_{pad}$.
The observed distribution is compared to a set of simulated ones
obtained with different test spectra,
to determine the spectrum that better reproduces the data.

%  \vspace*{0.5cm}

\section{The Crab Nebula Signal}

The data set used for this analysis contains all the events recorded
from 2007 November to 2013 February, with N$_{pad} \geq$ 20.
The total on-source time is 1.12 $\times$ 10$^4$ hours. 

For each source transit, the events are used to fill a set of nine 
12$^{\circ}\times$12$^{\circ}$ sky maps centered on the Crab Nebula position,
with a bin size of 0.1$^{\circ}\times$0.1$^{\circ}$ 
in right ascension and declination (``event maps'').
Each map corresponds to a defined N$_{pad}$ interval:
20-39, 40-59, 60-99, 100-199,
200-299, 300-499, 500-999, 1000-1999 and N$_{pad}\ge$ 2000.

To extract the excess of gamma rays, the cosmic ray
background has to be estimated and subtracted.
Using the {\em time swapping} method \citep{Ale92}, the shower data recorded 
in a time interval $\Delta t$ = 2-3 hours are used to evaluate 
the ``background maps'', i.e. the expected number of cosmic ray events  
in any location of the map for the given time interval.
This method assumes that during the interval $\Delta t$
the shape of the distribution of the 
arrival directions of cosmic rays in local coordinates does not change, 
while the overall rate could change due to atmospheric and detector effects.
The value of the time interval $\Delta$t 
is less than a few hours to minimize the
systematic effects due to the environmental parameters variations
that could change the distribution of the arrival directions.

The time swapping method is a sort of ``simulation'' based on real data:
for each detected event, 
$n_f$ "fake" events (with $n_f$ = 10) are generated by replacing the original
arrival time with new ones, randomly selected from an event buffer that
spans the time $\Delta$t of data taking. 
By changing the time, the fake events maintain the same declination of 
the original event,
but have a different right ascension. With these events a new sky
map (background map) is built, with a statistics $n_f$ times larger
than the event map in order to reduce the fluctuations. 
To avoid the inclusion of the source events in the background evaluation,
the showers inside a circular region around the source (with a radius 
related to the PSF and depending on N$_{pad}$) are excluded 
from the time swapping procedure. A correction on the number of swaps 
is applied to take into account the rejected events in the source region
\citep{fle04}.

Event and background maps are then smoothed 
according to the PSF corresponding to each N$_{pad}$ interval.
Finally, the smoothed background maps are subtracted to the
smoothed event maps, obtaining the ``excess maps'',
where for every bin the statistical significance of the excess is 
calculated as:

\begin{displaymath}
n_{\sigma} = \frac{N_E-N_B}{\sqrt{\delta N_E^2+ \delta N_B^2}} 
\end{displaymath}

with $N_E$ = $\Sigma_i$ $n_i$ $w_i$  and
$N_B$ = $\Sigma_i$ $b_i$ $w_i$ / $n_f$. 
In these expressions $n_i$ and $b_i$  are the number of events 
of the i$^{th}$ bin of the event map and background map,
respectively,
$w_i$ is a normalized weight, proportional to the value of the PSF 
at the angular distance of the i$^{th}$ bin, and
$n_f$ is the number of swappings.
The sum is over all the bins inside a radius $R_{max}$, chosen 
to contain the signal events and depending on the PSF.
Since the number of events per bin is large, 
the fluctuations follow the Gaussian
statistics, hence the errors on $N_E$ and $N_B$ are:
    $\delta N_E$ =  $\sqrt{\Sigma_i n_i w_i^2}$
and $\delta N_B$ = $\sqrt{\Sigma_i b_i w_i^2/n_f^2}$.

The number of gamma ray events from the source is:

\begin{displaymath}
 N_{\gamma} = \frac{N_E-N_B}{2 \pi \int_0^{R_{max}} w(r)^2 r dr}
\end{displaymath}

where $w(r)$ is the weight used in the smoothing procedure
calculated at the angular distance $r$ from the source position.

When adding all data,
an excess consistent with the Crab Nebula position is observed in each
of the 9 maps, with a total statistical significance of 21.1 standard deviations.
%------------------------------------------------------------------------
 \begin{figure}[t]
  \centering
  \includegraphics[width=0.45\textwidth]{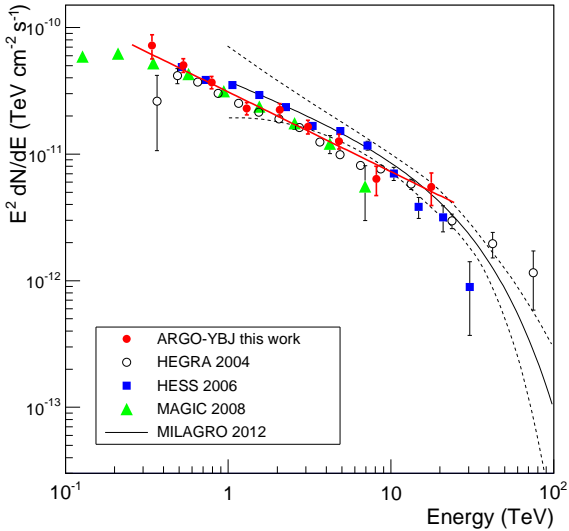}
  \caption{Crab Nebula differential energy spectrum multiplied by E$^2$ 
measured by ARGO-YBJ and other experiments. The thick solid line 
(red in the online version) represents
the best fit of the ARGO-YBJ data. The dotted lines delimit
the 1 sigma error band of the Milagro spectrum.}
  \label{spectrum}
  \vspace*{0.5cm}
 \end{figure}
%------------------------------------------------------------------------

The number of excess events are $\sim$3.3 $\times$ 10$^5$, corresponding to 
189 $\pm$ 16 day$^{-1}$, where a ``day'' means a source transit.
Table \ref{table2} gives the signal significance for each map and the 
corresponding event rates measured from the source. For comparison 
the background rates measured inside an angular window 
of 1$^{\circ}$ radius around the source are given.
Fig.~\ref{map} shows the total significance map.

Finally, the gamma ray signal
can be used to check the detector angular resolution since the Crab Nebula
angular size is small compared to the detector PSF. 
Fig.~\ref{ang_res} shows the distribution of the arrival directions of the
excess showers with respect to the source position, 
for N$_{pad} \ge$ 20, 100 and 500, compared to simulations. 
The agreement is excellent.

\begin{deluxetable*}{cccccc}
%\begin{table*}{cccccc}
\tabletypesize{\small}
\tablecaption{Summary of Crab Nebula data.}
\tablewidth{0pt}
\tablehead{
\colhead {N$_{pad}$} & 
\colhead{Photon rate} & 
\colhead{Background rate\tablenotemark{a}} &
\colhead{Significance} &
\colhead{E$_{med}$} &  
\colhead{Differential flux} 
\\
\colhead{} & 
\colhead{(events day$^{-1}$)} & 
\colhead{(events day$^{-1}$)} & 
\colhead{(s.d.)} & 
\colhead{(TeV)}  &
\colhead{(ph cm$^{-2}$ s$^{-1}$ TeV$^{-1}$)}  
}
\startdata
20-39     & 56.7 $\pm$ 12.2 & 1.3$\times$10$^4$ & 4.6 & 0.34  & (6.23 $\pm$ 1.34)$\times$10$^{-10}$\\ 
40-59     & 75.4 $\pm$ 9.2  & 1.1$\times$10$^4$ & 8.2 & 0.53  & (1.80 $\pm$ 0.21)$\times$10$^{-10}$\\ 
60-99     & 34.7 $\pm$ 3.9  & 4.2$\times$10$^3$ & 9.0 & 0.79  & (5.92 $\pm$ 0.66)$\times$10$^{-11}$\\ 
100-199   & 15.4 $\pm$ 1.7  & 1.9$\times$10$^3$ & 8.9 & 1.30  & (1.37 $\pm$ 0.15)$\times$10$^{-11}$\\ 
200-299   & 5.23 $\pm$ 0.61 & 4.9$\times$10$^2$ & 8.5 & 2.1   & (5.30 $\pm$ 0.63)$\times$10$^{-12}$\\ 
300-499   & 3.51 $\pm$ 0.44 & 3.8$\times$10$^2$ & 8.0 & 3.1   & (1.75 $\pm$ 0.22)$\times$10$^{-12}$\\ 
500-999   & 2.07 $\pm$ 0.27 & 2.4$\times$10$^2$ & 7.6 & 4.8   & (5.62 $\pm$ 0.74)$\times$10$^{-13}$\\ 
1000-1999 & 0.50 $\pm$ 0.13 & 87.4              & 3.8 & 8.1   & (1.00 $\pm$ 0.26)$\times$10$^{-13}$\\ 
$\ge$ 2000 & 0.23 $\pm$ 0.07 & 34.2              & 3.5 & 17.7  & (1.87 $\pm$ 0.54)$\times$10$^{-14}$\\
\enddata

\tablecomments{\\ $^a$ Average background rate within an angular distance of 1$^{\circ}$ from the source.} 

\label{table2}
%\end{table*}
\end{deluxetable*}

\section{Energy Spectrum}

The energy spectrum is evaluated by comparing the number of events detected
from the Crab Nebula in the previously defined N$_{pad}$ intervals 
to the expected number given by a simulation assuming a set of test spectra.
We consider the power law spectrum: 
$$
\frac{dN}{dE}(I_0,\alpha) = I_0 \biggl(\frac{E}{2 \; \mathrm{TeV}}\biggr)^{-\alpha}, 
$$
where the flux normalization I$_0$ and slope 
$\alpha$ are the parameters to be estimated with the fitting
procedure.

The fit is made by minimizing the value of $\chi^2$, evaluated for any
couple of parameters as:

%\begin{displaymath}
$$
\chi^2(I_0,\alpha) = \sum_{j=1,9} \frac{(N^j_{\gamma} - N^j_{MC}(I_0,\alpha))^2}{(\delta N^j_{\gamma})^2 + (\delta N^j_{MC})^2}  
$$
%\end{displaymath}

where $N^j_{\gamma}$ and  $N^j_{MC}$ are the number of events detected and expected,
respectively, in the $j^{th}$ N$_{pad}$ interval.

The obtained best-fit parameters are
$I_0$ = (5.2 $\pm$ 0.2) $\times$ 10$^{-12}$ photons cm$^{-2}$ s$^{-1}$ TeV$^{-1}$,
and $\alpha$ = -2.63 $\pm$ 0.05, 
with $\chi^2$ = 5.8 for 7 degrees of freedom, 
corresponding to a p-value $p$ = 0.56.
The integral flux above 
1 TeV is 1.97 $\times$ 10$^{-11}$ photons cm$^{-2}$ s$^{-1}$.
The flux at 1 TeV obtained in this work is 7$\%$ higher than
that reported in a previous ARGO-YBJ paper \citep{sky13}.
The difference is due to the correction of the event rates applied
in this work, to reduce environmental
and detector effects on the trigger rate, as described in Section 5.3.

Fig.~\ref{spectrum} shows the obtained spectrum compared 
with the results of other experiments.
The energy of each point
is the gamma ray median energy for the corresponding N$_{pad}$ interval. 
The values of energies and differential fluxes are given in Table \ref{table2}.
The spectrum is consistent with a constant slope from $\sim$300 GeV 
to $\sim$20 TeV and agrees rather well 
with the measurement by HEGRA and MAGIC, whereas the HESS and Milagro 
fluxes are about 20$\%$ higher in the $\sim$1-10 TeV energy range.

The data are less clear concerning a possible energy cutoff 
at higher energies.
MAGIC \citep{magic08} and HESS \citep{hess06} show a steepening below 20 TeV,
while
the HEGRA spectrum is harder and continues with a slight softening up to 
$\sim$75 TeV \citep{hegra04}.
A possible cutoff is also observed by Milagro at $\sim$30 TeV \citep{mila12}.
The limited statistics of our data at high energy does not allow to draw any conclusion 
about the spectral properties above 20 TeV.
Selecting events with  N$_{pad}\ge$ 3000 (whose median energy is 26 TeV
assuming a power law spectrum with index  $\alpha$ = -2.63) the statistical significance of
the signal is 0.75.

When fitting the data with a power law spectrum with an exponential cutoff:

$$
\frac{dN}{dE}(I_0,\alpha,E_{cut}) = I_0 \biggl(\frac{E}{2 \; \mathrm{TeV}}\biggr)^{-\alpha} \; \exp \; (-E/E_{cut}) 
$$

the obtained p-value is always smaller than without a cutoff, for any
value of $E_{cut}$.
For E$_{cut}$ = 14.3 TeV (the best-fit value obtained by HESS) 
the p-value is 0.13. 
We found that the p-value is larger than 10$\%$ for any value 
of E$_{cut}$ $>$ 12 TeV, indicating that
the presence of a cutoff above $\sim$10 TeV cannot be excluded
even if our data seems more consistent with a pure power law.

\subsection {Estimation of Systematic Errors}

The previous results can be affected by systematic errors of different origin.
In the following we discuss the possible sources of systematics,
evaluating their effects both on the flux normalization and the spectral slope.

1) {\it Energy scale.}
In our measurement the number of hit pads N$_{pad}$ is used as
an estimator of the primary energy. The relation between the primary energy
and  N$_{pad}$ is given by Monte Carlo simulations. Possible 
uncertainties and simplifications in the
simulation procedure (both in the shower development and the detector 
response) could produce an incorrect  N$_{pad}$ value and consequently 
an error in the energy scale. 

The energy scale reliability has been checked using the Moon shadow.
Due to the geomagnetic field, cosmic rays are deflected
according to their energy and the shadow that the Moon casts on the cosmic ray
flux is shifted with respect to the Moon position by an amount depending
on the energy.
The westward shift of the shadow 
has been measured for different N$_{pad}$ intervals and compared to
simulations.
From the analysis of the Moon data, we found that 
the total absolute energy scale error
is less than 13$\%$ in the proton energy range $\sim$1-30 TeV \citep{DiS11}. 
This estimate includes the uncertainties of the
cosmic ray elemental composition and the hadronic interaction model.

From this result, given a gamma ray spectrum with index $\alpha$ = -2.63, 
the corresponding systematic 
error in the flux normalization would be less than 22$\%$.

2) {\it Pointing error.}
Fitting the angular distribution of gamma rays around the Crab Nebula position
we found that the pointing error is less than 0.1$^{\circ}$.
A pointing error affects the measured gamma ray flux,
since the number of photons is obtained
by a smoothing procedure weighting the events with a PSF centered
at the source nominal position.
An incorrect position would produce a loss of signal. 
Since the PSF is narrower for events with large N$_{pad}$,
the loss is larger at high multiplicities, 
and generates a steepening of the spectrum.
According to our simulation, a pointing error of 0.1$^{\circ}$ would produce
a loss of signal ranging from 0.1$\%$ for  N$_{pad}$ = 20-39 to 6.0$\%$
for N$_{pad} >$2000. As a consequence, 
the spectral index would increase by 0.01 
and the flux normalization would decrease by 2$\%$.

3) {\it Background evaluation.}
Our measurement is based on a very precise evaluation of the background.
As explained in Section 3, the number of gamma rays is 
given by the difference between the number 
of events detected in the event map (that contains the 
source events plus the cosmic ray
background) and the number of background events 
estimated with the time swapping method.
Since the ratio between the number of gamma rays and the number
of background events is very small (ranging
from $\sim$3 $\times$ 10$^{-4}$ for 
N$_{pad}$ = 20-39 up to $\sim$4 $\times$ 10$^{-2}$ for N$_{pad}$ $>$ 300),
even a small systematic error in the background evaluation could produce
a big error in the source flux.

Possible sources of systematics are: 
{\it a)} the presence of cosmic ray excess regions due to
the medium scale anisotropy, as those reported in \citep{aniso13}, close to the
source,
{\it b)} changes in atmospheric conditions able to modify 
the background distribution in local coordinates in less than 2-3 hours, 
and {\it c)} the detector malfunctioning.

Such effects, when present, could 
generate extended regions in the signal map 
with evident excesses or deficits, in some cases involving the whole map.
Instead, an accurate evaluation of the background
produces a map with all the bin contents consistent with zero 
except at the position of real sources.
 
Concerning the medium scale anisotropy, we adopted 
a particular procedure to correct the background systematics
in the sky regions coincident or adjacent to cosmic ray excesses
\citep{sky13}. This correction is not necessary
in the Crab Nebula region. 
 
Concerning points $b$ and $c$, it has to be specified that the 
maps are built with data sets of 2-3 hours and individually checked.
When a map shows significant anomalies, the corresponding
dataset is rejected, so only ``good maps'' are combined to build
to the ``total maps''.

To test the background reliability of the total maps, 
we use the regions that are not
involved in the Crab Nebula emission, i.e., the bins 
with an angular distance
from the source larger than a minimum value, depending on the PSF.
From these ``out-source'' regions we expect no significant
excess, since they do not include any other known gamma ray
source with a flux above the ARGO-YBJ sensitivity.
For any of the nine maps we have evaluated the distribution of the
excesses in the out-source region bins (before smoothing). 
We found that all the 
distributions are well described by Gauss functions with mean values
consistent with zero and r.m.s. consistent with unit.

Adding all the nine maps together, 
the total number of events detected from the out-source regions 
is 1.18 $\times$ 10$^9$. This value differs from the
corresponding estimated background by -9.3 $\times$ 10$^3$ events, 
corresponding to -0.3 standard deviations.
Since there is no significant excess or deficit of events,
we can calculate the upper limit of the systematic error in the
out-source region. We found that the relative error in the
background value is less than 3.7 $\times$ 10$^{-5}$ at 90$\%$ confidence level.

We can reasonably assume that a similar systematic error involves the
region of the map containing the source signal, and that
all the nine maps have comparable systematic errors.
Based on these assumptions, we can evaluate the effects of such an error 
on the signal, which are obviously more relevant for the maps 
in which the signal-to-background ratio is smaller. 
We found that the error in the photon number is $<$13$\%$ for N$_{pad}$ = 20-39,  
$<$1$\%$ for N$_{pad}$ = 100-199, and $<$0.01$\%$ for  N$_{pad}$ $\ge$ 1000.

According to these values, the corresponding 
systematic error in the spectrum flux normalization would be less than 2$\%$, 
and the error in the spectral index would be less than 0.05.

4) {\it Event rate variations.}
Studying the rate of cosmic ray showers over five years, we observed 
variations on timescales from hours to months
up to 10$\%$ with respect to the mean value.
These variations are mostly due to:
{\it a)} variation of atmospheric pressure and temperature that modify
the showers propagation in the atmosphere,
{\it b)} variation of the detector efficiency
due to changes of the local temperature and pressure,
{\it c)} aging of the detector.

Gamma rays are assumed to be subject to similar variations.
To study the stability of the Crab Nebula flux, we 
corrected the rate of the events observed from the source
using the cosmic ray rate as a normalization factor (see Section 5.3). 
However, an absolute normalization cannot be performed.
The Monte carlo simulations refer to a fixed atmospheric condition
and a given detection efficiency, that cannot exactly reproduce
the average effect over several years of different conditions.
Considering the amount of the observed rate variation, 
a reasonable estimation indicates a possible systematic error in the flux
smaller than 4$\%$. 

{\it Total systematic error.}
Adding all these contributions {\it linearly}, we conservatively estimate
the total systematic error to be less than 30$\%$ for the flux normalization
and 0.06 for the spectral index.

\section{Crab Nebula Light Curve}

%------------------------------------------------------------------------
 \begin{figure}[t]
  \centering
  \includegraphics[width=0.45\textwidth]{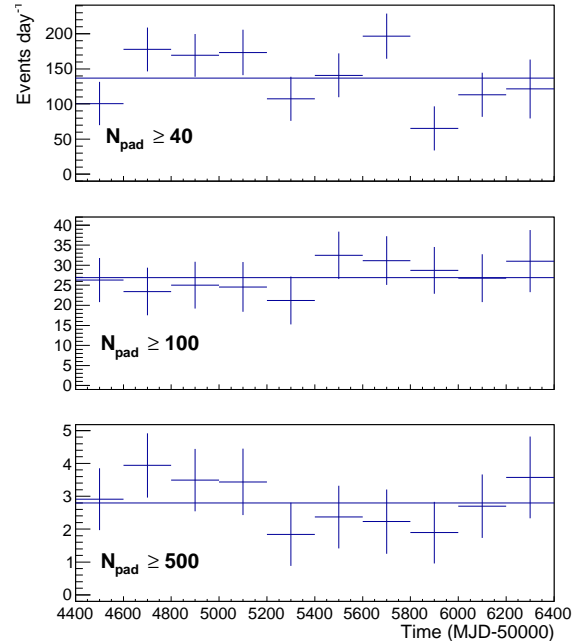}
  \caption{Rate of events detected from the Crab Nebula 
for different N$_{pad}$ thresholds as a function of time. 
The bin width is 200 days. 
The lines represent the average values.}
  \label{rate}
  \vspace*{0.5cm}
 \end{figure}

%------------------------------------------------------------------------

To study the stability of the Crab Nebula emission,
we consider the events with N$_{pad}\ge$ 40.
Our total dataset consists of 1851 days, with an average observation
time of 6.0 hours per day. 
The average rate of events with N$_{pad}\ge$ 40
is 137 $\pm$ 10 day$^{-1}$.

Figure \ref{rate} shows the observed rate for events with N$_{pad}\ge$ 40, 
100 and 500, as a function of the Julian date, in bins of 200 days.
The median energies corresponding to these  N$_{pad}$ thresholds are
0.76, 1.8 and 5.1 TeV, respectively.
The signal appears stable during five years for any threshold 
within the statistical fluctuations.
Assuming a constant rate, the obtained $\chi^2$ are 15.7, 3.27 and 5.17
(with 9 d.o.f.) for  N$_{pad}\ge$ 40, 100 and 500.
The corresponding p-values are 0.073, 0.95 and 0.82, 
respectively.

A six-year monitoring of the Crab Nebula was previously performed 
by the Tibet-III air shower array from 1999 to 2005, at energies $\sim$3 TeV, 
with a sensitivity 3-4 times lower than that of ARGO-YBJ,
reporting a yearly flux consistent with a steady emission \citep{tibet09}.

\subsection{Search for Flares}

To make a ``blind'' search for short time rate variations,  
we consider all the time intervals of duration $\Delta t$ 
ranging from 1 to 15 days, starting from every observation day. 
This time range has been chosen on the basis of the duration of the flares
observed in the GeV energy region. 
For each interval we compare the observed rate of Crab events 
with the average rate and evaluate the significance of the excess as:
$\sigma _i$ = ($R_i-R_m$) / $\delta (R_i-R_m)$, 
where $R_i$ is the counting rate in the i-th interval,
$R_m$ is the average counting rate, and
$\delta (R_i-R_m)$ is the statistical error of the difference $R_i-R_m$.
Note that the values of $\sigma _i$ are not independent, 
since the time intervals overlap.

\begin{figure}[t]
  \centering
  \includegraphics[width=0.45\textwidth]{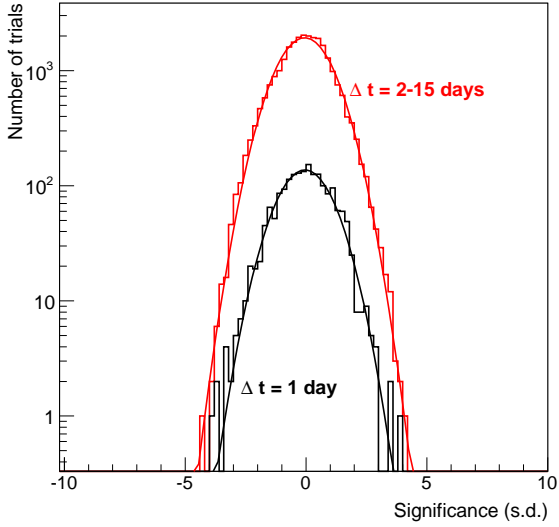}
  \caption{Search for flares: 
distribution of the excesses from the Crab Nebula around 
the average value, in units of standard deviations, for different flare
durations $\Delta t$.
}
 \vspace*{0.5cm}
 \label{gauss}
 \end{figure}

Fig.~\ref{gauss} shows the distributions of $\sigma _i$
for $\Delta t$ = 1 day and $\Delta t$ = 2-15 days, for N$_{pad}\ge$ 40.
The total number of intervals is 1851 for $\Delta t$ = 1 and 25911  for 
$\Delta t$ = 2-15 days.
The distributions can be fitted by a Gauss function with mean value 
$m$ = -0.04 $\pm$ 0.03 and r.m.s.= 1.05 $\pm$ 0.02 for $\Delta t$ = 1 day, and
$m$ = -0.06 $\pm$ 0.01 and r.m.s.= 1.061 $\pm$ 0.005 for $\Delta t$ = 2-15 days.
The root mean square 
values indicate rate variations slightly larger than what expected 
by statistical fluctuations. However, no significant excess is observed for any 
of the considered time intervals.

Given the ARGO-YBJ sensitivity, a flare would produce a 5 s.d. signal
(pre-trial) if the flux exceeds the average value 
by a factor $f$ $\sim$ 10 / $\sqrt{\Delta t (\mathrm{days})}$.

\subsection{Correlation with Fermi-LAT Data}

%------------------------------------------------------------------------

%------------------------------------------------------------------------

To reduce the number of trials in the search for possible flares,
we can limit our analysis to the days in which a flare was observed by
satellite instruments at lower energies.
We consider the Fermi-LAT daily light curve at energy E $>$ 100 MeV 
from 2008 August to 2013 February, obtained through 
the analysis of the scientific Fermi data publicly available  
at the Fermi Science Support 
Center\footnote{http://fermi.gsfc.nasa.gov/ssc/data/access/}.

\begin{figure}[t]
  \centering
  \includegraphics[width=0.45\textwidth]{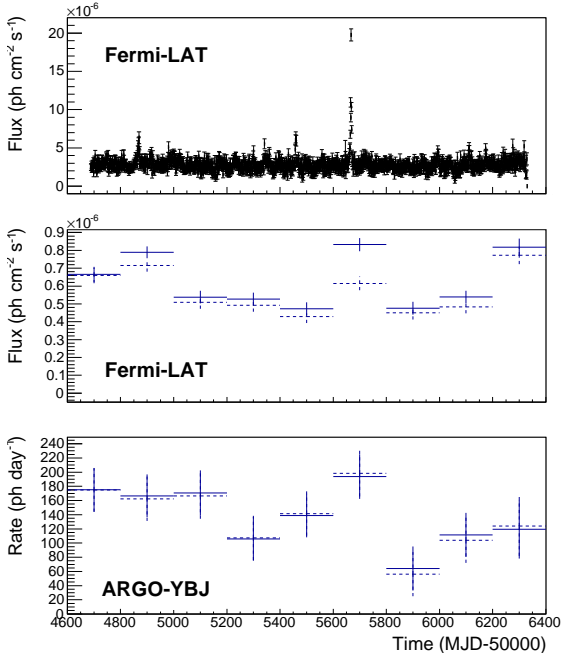}
  \caption{Panel 1: daily light curve of the Crab Nebula and pulsar
            by Fermi-LAT; 
           Panel 2: light curve of the Crab Nebula (pulsar subtracted) 
           in bins of 200 days, by Fermi-LAT; 
           Panel 3: light curve of the Crab Nebula by ARGO-YBJ  for events with N$_{pad}\ge$ 40.
           The dashed line in panels 2 and 3 
           has been obtained excluding the days with flares.}
  \label{correl}
  \vspace*{0.5cm}
 \end{figure}

The first panel of Figure~\ref{correl} shows the daily light curve, 
representing the sum of the nebula and pulsar fluxes.
The average flux is (2.66 $\pm$ 0.01) $\times$ 10$^{-6}$ photons cm$^{-2}$ s$^{-1}$.
Also excluding the days with flares, the rate
is variable, with modulations on timescales of weeks and months.

%-------------------------------------------------------------------

\begin{deluxetable*}{ccccccc}
\tabletypesize{\small}
\tablecaption{ARGO-YBJ results during the 
three largest Crab Nebula flares observed by Fermi-LAT.}
\tablewidth{0pt}
\tablehead{
\colhead {} & 
\colhead {$\Delta$t} & 
\colhead {Max.Fermi flux} &
\colhead {ARGO-YBJ} &
\colhead {ARGO rate} & 
\colhead {ARGO rate} & 
\colhead {ARGO rate} \\ 
\colhead {} & 
\colhead {(days)} & 
\colhead {(ph cm$^{-2}$ s$^{-1}$)} & 
\colhead {Observation} &
\colhead {N$_{pad} \ge$40} & 
\colhead {N$_{pad} \ge$100} &
\colhead {N$_{pad} \ge$500} \\
\colhead {} &
\colhead {} &
\colhead {} &
\colhead {time (hr)} &
\colhead {(ph day$^{-1}$)} &
\colhead {(ph day$^{-1}$)} &
\colhead {(ph day$^{-1}$)}
}
\startdata
Flare 1\tablenotemark{a}&  8  & 6.3$\pm$0.8$\times$10$^{-6}$ & 49.6 & 142 $\pm$ 151 & 21 $\pm$ 28 & 2.5  $\pm$ 4.6  \\
Flare 2\tablenotemark{b}&  5  & 6.4$\pm$0.8$\times$10$^{-6}$ & 31.5 & 265 $\pm$ 190 & 58 $\pm$ 36 & -3.8 $\pm$ 5.7  \\
Flare 3\tablenotemark{c}&  9  &19.8$\pm$0.8$\times$10$^{-6}$ & 58.0 & 228 $\pm$ 144 & 51 $\pm$ 27 & 2.9  $\pm$ 4.4  \\
Sum of 3 flares         & 22  &      & 139            & 205 $\pm$ 91 & 41  $\pm$ 17 & 1.3  $\pm$ 2.8  \\
All ARGO data            &     &      &                & 137 $\pm$ 10 & 27  $\pm$ 2  & 2.8  $\pm$ 0.3  \\
\enddata

\tablecomments{\\ $^a$ Start time MJD 54864 (2009 February 02) \\   $^b$ Start time MJD 55457 (2010 September 18) \\
$^c$ Start time MJD 55662 (2011 April 11)}

\label{tabflare}
\end{deluxetable*}

%------------------------------------------------------------------

%------------------------------------------------------------------------

%------------------------------------------------------------------------

First we consider the three largest Fermi flares, which 
occurred in 2009 February,
2010 September and 2011 April \citep{fermi1,fermi2}.
To define the time boundaries and the duration of these flares 
we select the days in which the Fermi flux is higher than
4 $\times$ 10$^{-6}$ photons cm$^{-2}$ s$^{-1}$.
The dates and the duration of the three flares are given in Table
\ref{tabflare}.
The counting rates from the Crab Nebula measured by ARGO-YBJ
with events with  N$_{pad}\ge$ 40
during the flares are compared with the average rate
of 137 $\pm$ 10 day$^{-1}$.
In all cases, the rates are slightly higher than the average value, 
but consistent with it within statistical errors (see Table \ref{tabflare}).
Summing the three flares the average rate is 205 $\pm$ 91 day$^{-1}$.
Table \ref{tabflare} also shows the results concerning 
the events with N$_{pad}\ge$ 100 and 500. 
No significant excess is present in this case, either. 

Our preliminary analysis reported in \citep{argo10} showed a 4 standard deviations excess 
observed in the time interval from 2010 September 17 to 22 
from a direction consistent with the Crab Nebula. 
However, removing the contribution of the steady flux and taking into account the number of trials, 
the post-trial significance was about two standard deviations. 
A further excess with a similar post-trial statistical significance was observed during the 
2011 April flare \citep{ver_scineghe}.
In the present work, based on a better shower reconstruction and the event selection described in Section 2.2, 
the significance of the Crab Nebula signal integrated over 5 years increases by about 15$\%$ with respect to the 
old analysis, but the signal observed during the Fermi flares decreases. 
The flux measured during the flares appears slightly higher than what 
would be expected from the steady emission, 
but consistent with it within one standard deviation. Both our previous analysis and the current one 
hint at a possible flux enhancement during the flares, but the reduced significance 
prevents us from drawing a definitive conclusion.

To extend the search for flares to the whole observation time, and not limit the
analysis to the largest flares, we selected the Fermi data according to the
measured daily flux and checked the corresponding ARGO-YBJ event rate.
Table \ref{tabdays} reports the ARGO-YBJ rates for different
levels of the Fermi flux and different N$_{pad}$ thresholds. 
The rates are consistent with the average rate for any Fermi flux level.
In particular, the ARGO-YBJ rate (for N$_{pad}\ge$ 40) 
measured in the 62 days in which the Fermi flux 
exceeds 4 $\times$ 10$^{-6}$ ph cm$^{-2}$ s$^{-1}$, is 190 $\pm$ 55 day$^{-1}$ 
i.e. 1.4 $\pm$ 0.4 times higher than the average rate.

\begin{deluxetable*}{ccccc}
\tabletypesize{\small}
\tablecaption{ARGO-YBJ photon rate for different flux levels measured by Fermi.}
\tablewidth{0pt}
\tablehead{
\colhead {Fermi flux} & 
\colhead {Number of days} & 
\colhead {ARGO rate} &
\colhead {ARGO rate} &
\colhead {ARGO rate}  \\
\colhead{( 10$^{-6}$ ph cm$^{-2}$ s$^{-1}$)} &
\colhead{} &
\colhead {N$_{pad} \ge$40} & 
\colhead {N$_{pad} \ge$100} &
\colhead {N$_{pad} \ge$500} \\
\colhead{} &
\colhead{} &
\colhead{(ph day$^{-1}$)} &
\colhead{(ph day$^{-1}$)} &
\colhead{(ph day$^{-1}$)}
}
\startdata
$<$ 2.0    & 175 & 197 $\pm$ 34  & 26 $\pm$ 6  & 3.1 $\pm$ 1.0 \\ 
2.0-3.0    & 915 & 118 $\pm$ 15  & 27 $\pm$ 3  & 2.8 $\pm$ 0.4 \\ 
3.0-4.0    & 435 & 148 $\pm$ 21  & 28 $\pm$ 4  & 3.0 $\pm$ 4.3 \\ 
4.0-5.0    & 46  & 188 $\pm$ 64  & 33 $\pm$ 12 & 1.9 $\pm$ 2.0 \\ 
$>$ 5.0    & 16  & 198 $\pm$ 107 & 54 $\pm$ 20 & 0.2 $\pm$ 3.2 \\
All ARGO data   &     & 137 $\pm$ 10  & 27  $\pm$  2  & 2.8 $\pm$ 0.3  \\
\enddata
\label{tabdays}
\end{deluxetable*}

%-------------------------------------------------------------------

Finally, to study a possible correlation on timescales of months or years,
we compare the light curves of the two detectors
over the common observing time ($\sim$ 4.5 years),
dividing the data into bins of 200 days.
The bin width is chosen  in order to have a significant 
signal in the ARGO-YBJ data (about 7 s.d.). 

Since the flux measured by Fermi is the sum of the nebula 
and pulsar contributions, and since
the pulsar flux $F_P$ averaged over the pulsation period is
also stable during flares 
($F_P$ = (2.04 $\pm$ 0.01) $\times$ 10$^{-6}$ photons cm$^{-2}$ s$^{-1}$
for E $>$100 MeV \citep{fermi2}),
the flux of the pulsar has been subtracted.
The obtained average nebula flux is (6.2 $\pm$ 0.1) 
$\times$ 10$^{-7}$ photons cm$^{-2}$ s$^{-1}$.
The nebula flux shows variations up to $\sim$30$\%$ of the average flux,
with $\chi^2$ = 126, for 8 d.o.f. (see the second panel of Fig.~\ref{correl}).
The large variations are not only due to flares.
In the same figure the dashed curve shows the flux
obtained excluding the 62 ``flaring days''. In this case 
the average value is (5.6 $\pm$ 0.1) $\times$ 10$^{-7}$ photons
 cm$^{-2}$ s$^{-1}$, with $\chi^2$ = 80. 

The lower panel of Fig.~\ref{correl} shows the corresponding ARGO-YBJ data 
for N$_{pad}\ge$ 40. 
The average rate is 139.3 $\pm$ 10.6 events day$^{-1}$ ($\chi^2$ = 14.2 for 8 
d.o.f., p-value $p$ = 0.077).
Even if the ARGO-YBJ rate variations are consistent 
with statistical fluctuations,
the Fermi and ARGO-YBJ data seems to follow a similar trend.
The ARGO-YBJ rate appears higher in the ``hot'' Fermi periods.
The dashed curve is obtained after the exclusion of the flaring days.

\begin{figure}[tbp]
  \centering
  \includegraphics[width=0.45\textwidth]{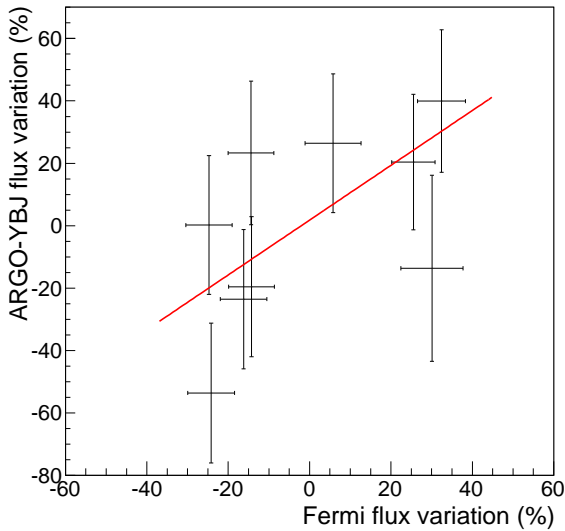}
  \caption{Percentage variation of the Crab Nebula flux 
with respect to the average value: ARGO-YBJ vs Fermi-LAT data. 
The straight line is the best fit curve.}
  \label{correl2}
  \vspace*{0.5cm}
 \end{figure}

Fig.~\ref{correl2} shows the ARGO-YBJ percentage rate variation  
with respect to the mean value ($\Delta F_{ARGO}$)
as a function of the corresponding variation
of the Fermi rate  ($\Delta F_{Fermi}$), for the 9 bins of the light curve.
The Pearson correlation coefficient between the two data sets
is $r$ = 0.56 $\pm$ 0.22.
The quoted error for $r$ is the root mean square of the
distribution of the correlation coefficients obtained by simulating
the fluctuations of the counting rates of each bin, according to their 
statistical errors.
Fitting the 9 points with the function
$\Delta F_{ARGO}$ = $a$ $\Delta F_{Fermi}$ + $b$,
the values of the best-fit parameters 
are $a$ = 0.88 $\pm$ 0.37 and $b$ = 0.018 $\pm$ 0.079, 
with $\chi^2$ = 8.3 for 7 d.o.f.
Discarding the 62 ``flaring'' days the correlation coefficient 
becomes 
$r$ = 0.45 $\pm$ 0.23 and the parameters of the linear fit are 
$a$ = 0.96 $\pm$ 0.45 and $b$ = 0.018 $\pm$ 0.082, with $\chi^2$ = 10.4. 

The same analysis has been performed using a different bin width, ranging from
10 to 450 days. The corresponding correlation coefficient steadily increases
from $r$ = 0.10 $\pm$ 0.06 (10 days) to $r$ = 0.59 $\pm$ 0.23 (450 days).
It has to be noted, however, that when using a small bin width, 
the ARGO-YBJ signal
is not significant enough to search for a correlation unless the flux
variations are very large. In 10 days, for example, the average ARGO-YBJ
signal is 137 $\pm$ 135  events day$^{-1}$. The statistical fluctuations would
hide a possible flux variation, unless the flux becomes more than a factor
of 4-5 higher than the average.

The above results refer to events with  N$_{pad}\ge$ 40.
The correlation coefficient is lower when selecting more energetic events: 
using a bin width of 200 days, for  N$_{pad}\ge$ 100, $r$ = 0.19 $\pm$ 0.31; 
and for  N$_{pad}\ge$ 500, $r$ = 0.46 $\pm$ 0.28.

\subsection{Stability of the ARGO-YBJ Data}

When studying the time evolution of a signal over
several years, a discussion on the possible causes of 
detector instabilities
is mandatory, to exclude systematic effects that could produce
artificial rate variations.
Since the measured number of events from the source $N_S = N_E-N_B$ 
is the difference between the number of events
$N_E$ detected in the source map and the number of background events $N_B$
estimated with the time swapping method, one must separately analyze the
stability of the different contributions.

1) A loss of signal events $N_S$ could be produced by variations
of the pointing accuracy.
Studying the Moon shadow month by month, we have verified that 
the pointing is stable within 0.1 deg \citep{DiS11}. 
Given the moderate angular resolution
for events with N$_{pad}\ge$ 40, such a value could produce
signal fluctuations of less than 2$\%$.

2) A worsening of the detector angular resolution (due to an increase
of the time resolution of RPCs occurring at particularly low temperatures)
could produce a loss of signal events $N_S$. 
A broadening of the PSF 
would also cause a decrease of the Moon shadow signal.
That, however, is found to be stable within statistical fluctuations.
 
3) Atmospheric pressure and temperature variations can affect
the RPC detection efficiency, which can also be altered by some RPC 
not working properly or by aging effects.

4) Pressure and temperature produce changes in the shower rate
of the order of a few percent due to the different conditions in which
the showers propagate in the atmosphere.

The two latter effects modify $N_S$, $N_E$ and $N_B$ 
by about the same factor (neglecting the different behavior
of cosmic ray and gamma ray showers, that in this contest 
can be considered a second-order effect). 
This allows the use of $N_B$ to correct the Crab rate,
multiplying the Crab rate observed in a given time interval  
by the correction factor $f_c$ = $B_m$/$B$, where $B_m$ is the 
average background rate and $B$ is the 
background rate in that interval.
The light curve in Figure \ref{rate} has been corrected according 
to this method, with $f_c$ ranging from 0.91 to 1.07. 

5) Further possible systematics could be an incorrect evaluation
of the background $N_B$. In Section 4 we evaluated 
the accuracy of the background for the total source signal.
To check the accuracy of the background along the years, 
we can use the same out-source regions previously defined.
For events with N$_{pad} \ge$ 40, the out-source  
light curve in 200-day bins has a 
mean value of -7.9 $\pm$ 19.0 events day$^{-1}$ and
a $\chi^2$ = 10.2 for 9 d.o.f.,
corresponding to a p-value $p$ = 0.67.
According to these results the background 
is stable and should not introduce any systematic effect on 
the rate of the Crab signal.

%---------------------------------------------------------------

%--------------------------------------------------------------

\section{Summary and Conclusions}

The ARGO-YBJ events recorded over five years have been analyzed
to evaluate the Crab Nebula spectrum and  
study the temporal behavior of the gamma ray emission.
Using the events with N$_{pad}\ge$ 20, the statistical significance 
of the gamma ray signal 
reaches more than 21 standard deviations,
and the observed photon rate is 189 $\pm$ 16 day$^{-1}$.
The event angular distributions around the source are well described by the
PSFs obtained by simulations.

The source spectrum extends over nearly 2 decades in energy 
and five decades in flux.
The spectral shape is consistent with a power law behavior in the 
range 0.3-20 TeV with a spectral index -2.63 $\pm$ 0.05.
An exponential cutoff would be consistent with our data
in case of a cutoff energy higher than 12 TeV at 90$\%$ confidence level. 

The study of the Crab Nebula light curve has been carried out 
to check the stability of the flux over years and to search
for possible flares on the timescale of days.
All the known sources of rate instabilities have been examined
and the effects corrected.

Concerning flares, 
a blind search for flux increases of duration between 1 and 15 
days shows no significant excess.
The average rate of events with N$_{pad}\ge$ 40
measured by ARGO-YBJ during  
the three most powerful flares detected by Fermi-LAT
(in 2009 February, 2010 September and 2011 April)
is 205 $\pm$ 91 day$^{-1}$, 
which is consistent with the average value of 137 $\pm$ 10 day$^{-1}$.

The five year ARGO-YBJ light curve 
with a binning of 200 days is consistent with a constant flux 
with a probability of 0.07. 
A correlation analysis with the corresponding Fermi-LAT data
gives a Pearson correlation coefficient $r$ = 0.56 $\pm$ 0.22.
The small statistical significance of these results does not allow the claim 
for a flux variability correlated with the observations at lower energies.
If such a correlation was due to a real astrophysical phenomenon, 
the found regression coefficient $a$ = 0.88 $\pm$ 0.37 would imply 
a similar percentage variation in Fermi and ARGO-YBJ rates,
suggesting a similar behavior
of the gamma ray emission at energies $\sim$ 100 MeV and $\sim$1 TeV.

So far, no variation of the Crab Nebula flux at TeV energies
has been reported by any detector.
Assuming the flares observed by AGILE and Fermi
due to synchrotron radiation
from a population of electrons accelerated up to 10$^{15}$ eV,
the Inverse Compton emission associated with this population
would occur in the Klein-Nishina regime and
would produce gamma rays of energy approximately equal 
to that of the electrons.
Such a flux would not be detectable by any of the existing gamma ray
experiments.
With these assumptions, a TeV excess could hardly be intepreted as
IC emission associated with the synchrotron radiation observed at lower energies,
and would require a completely new interpretation.

\acknowledgments

This work is supported in China by NSFC (No. 10120130794), the Chinese Ministry
of Science and Technology, the Chinese Academy of Science, the Key Laboratory
of Particle Astrophysics, CAS, and in Italy by the Istituto Nazionale di
Fisica Nucleare (INFN).

We also acknowledge the essential support of W.Y. Chen, G. Yang, X.F. Yuan, 
C.Y. Zhao, R. Assiro, B. Biondo, S. Bricola, F. Budano, A. Corvaglia,
B. D'Aquino, R. Esposito, A. Innocente, A. Mangano, E. Pastori, C. Pinto,
E. Reali, F. Taurino, and A. Zerbini, in the installation, debugging, and
maintenance of the detector.

\end{document}